\documentclass[prb,twocolumn,preprintnumbers,amsmath,amssymb]{revtex4}
\usepackage{color}
\usepackage{graphicx}
\usepackage{dcolumn}
\usepackage{verbatim}
\newcommand{\unit}[1]{\ensuremath{\, \mathrm{#1}}}

\begin{document}

\title{Prediction Of A Multi-Center Bonded Solid Boron Hydride for Hydrogen Storage}

\author{Tesfaye A. Abtew}
\author{Bi-ching Shih}
\author{Pratibha Dev}
\affiliation{Department of Physics, University at Buffalo, State University of New York, Buffalo, New York 14260, USA}

\author{Vincent H. Crespi}
\affiliation{Department of Physics and Department of Materials Science and Engineering, The Pennsylvania State University, 104 Davey Lab, University Park, Pennsylvania, 16802, USA}

\author{Peihong Zhang}
\email{pzhang3@buffalo.edu}

\affiliation{Department of Physics, University at Buffalo, State University of New York, Buffalo, New York 14260, USA}
\date{\today}

\maketitle

{\bf
An ideal material for on-board hydrogen storage
must release hydrogen at practical temperature and pressure and also 
regenerate efficiently under similarly gentle conditions.
Therefore, thermodynamically, the hydride material must lie within a narrow range near the hydrogenation/dehydrogenation phase boundary. Materials
involving only conventional bonding mechanisms are unlikely to meet these requirements.
In contrast, materials containing certain frustrated bonding
are designed to be on the verge of frustration-induced phase transition, and 
they may be better suited for hydrogen storage.
Here we propose a novel layered solid boron hydride and show its potential for hydrogen storage. 
The absence of soft phonon modes confirms the dynamical stability of the structure. Charging the structure significantly softens hydrogen-related phonon modes. Boron-related phonons, in contrast, are either hardened or not significantly affected by electron doping. These results suggest that electrochemical charging may facilitate hydrogen release while the underlying boron network remains intact for subsequent rehydrogenation.
}

Hydrogen fuel cells are attractive for their very high efficiency. Beyond
the technology of the cells themselves, these systems present two major
challenges to current technologies: economic large-scale hydrogen production
and practical small-scale on-board hydrogen storage~\cite{crabtree}. The storage problem in
particular demands fundamentally new transformative science and engineering,
since the standard modes of chemical and physical interaction known for
hydrogen appear to be grossly insufficient to meet the needs.
All current hydrogen storage technologies (e.g. cryogenic liquid, high-pressure
gas cells, low-temperature adsorbates, metal hydrides, and chemical storage)
suffer from one or more major deficiencies in cost, thermodynamic efficiency, volumetric
capacity, kinetics, gravimetric density, long-term storage and reversibility~\cite{schlapbach}. These deficiencies are so strong that a clear engineering
path to a practical storage system has not yet been envisioned. 
For example, transition metal hydrides usually have desired hydrogen release 
kinetics but fall short of the target storage capacity. 
Physisorption storage schemes, on the other hand, suffer from thermodynamic issues of storage capacity.
Complex hydrides formed from the restricted class of light metals 
(e.g. MAlH$_4$~\cite{bogdanovic97} and MBH$_4$~\cite{zuttel03-1}, where M denotes light metal such as Na or Li) and amide materials~\cite{chen}
suffer from issues of kinetics and variable desorption energies for successive hydrogens release. 
Systems based on amino-boranes (e.g. NH$_3$BH$_3$~\cite{fakioglu,gutowaska} and related compounds) 
have a fine balance of chemical energies that enable hydrogen release under relatively 
gentle conditions. However, successive hydrogens are progressively harder to remove, and 
the final dehydrogenated compound is a refractory ceramic that is not amenable to reversible rehydrogenation. 

It is interesting that all three major classes of hydrogen storage materials investigated 
so far (i.e.,  low temperature adsorbates, metal hydrides, and chemical storage) overlaps 
with regards to only boron. Examples include physical adsorption systems with boron-doped 
carbons~\cite{zhu06,sankaran05,kim06}, chemical systems using ammonia borane derivatives 
and light metal borohydrides. Boron is distinguished from the remainder of the periodic 
table by an {\it electronic frustration}. As a result, boron-based materials
are well-known for their rich and highly unconventional chemistry.
Having only three valence electrons, boron cannot 
form a sufficient number of traditional covalent bonds to reach a close-shell
configuration. Being a small, tight first-row p-block
element, it is insufficiently electropositive to participate strongly in
traditional metallic bonding, yet also insufficiently electronegative to 
be dominated by highly stable ionic configurations. These properties force
boron to assume unusual multi-center bonding configurations in many of its compounds.
One of the most famous examples, and a distinct feature of electron deficiency, 
is the three-center two-electron (3c2e) found in many boranes.
Chemically frustrated boron, when properly deployed in a supporting 
chemical framework, provides an intriguing opportunity to attain an 
intermediate bonding with hydrogen.
In this paper, we exploit the rich chemistry of boron and hydrogen
to propose a novel solid boron hydride structure which contains 
intermediate strength 3c2e-like multi-center B$-$H bonds and 
investigate its potential for on-board hydrogen storage.

\section*{Model}
\label{model}

Although the rich boron chemistry explored to date has been dominated by three dimensional
cage structures~\cite{perkins,hubert,scwacki}, computational and experimental studies~\cite{alexandrova,zhai,Lau} 
have begun to uncover a latent 
planar aspect in boron-rich materials. 
In the past decades, ab-initio results of different planar and quasi-planar structures of boron clusters have been proposed~\cite{boustani95,ricca97,zhai02}. Though these clusters were studied experimentally, understanding of their structural and electronic properties has been a challenge. Most recently, photoelectron spectroscopy (PES) technique~\cite{zhai03,sergeeva08} is used to unravel better structural information of boron clusters.
Using a combination of PES technique and ab-initio calculations, a stable planar structure of boron clusters containing ten to fifteen atoms~\cite{zhai03,sergeeva08} are shown to exist. 
Furthermore, a new class of planar boron sheets are predicted to exist through a self-doping mechanism~\cite{Ismail,lau08}. The stability of these sheets are explained with a bonding mechanism which arises from the competition of a  two- and three-center bonding. 

\begin{figure}[htp]
\includegraphics[width=0.45\textwidth]{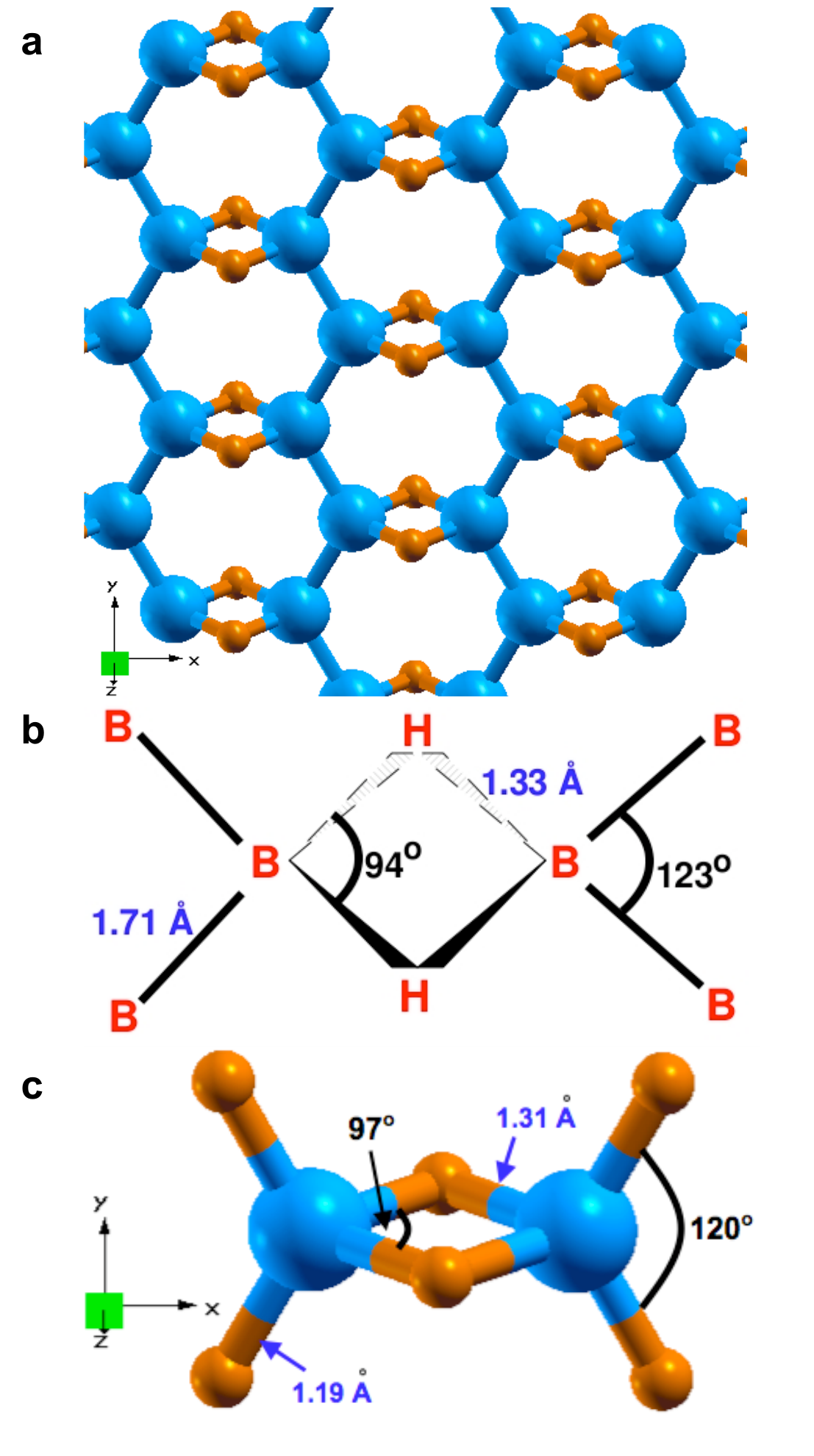}
{\bf \caption{Structure of single-layer B$_2$H$_2$: a, The relaxed structure. b, The corresponding bond angles and bond lengths. c, The structure of diborane. Boron is blue and hydrogen is orange.}
\label{B2H2-I-struc}}
\end{figure}

Pure hexagonal boron layer, although not stable alone due to electron deficiency, 
can be stabilized by intercalating electron donors such as Mg and Al to 
form MgB$_2$ and AlB$_2$, (and also TaB$_2$, HfB$_2$, ZrB$_2$, and TiB$_2$).
This hexagonal boron sheet has a much greater structural flexibility than does graphite:
for example, the in-plane lattice constant expands from $3.01\unit{\AA}$ in AlB$_2$ 
to $3.17\unit{\AA}$ in ZrB$_2$. 
Very recently, the possibility of using planar boron sheets with possible alkali metals as a template for hydrogen storage has been investigated~\cite{suleyman09}.  
In this work, we propose that the electron donor can also be hydrogen atoms.
In fact, nature finds a way to stabilize the electron-deficient 
molecule diborane through the formation of 3c2e bonds between
boron and hydrogen. It is plausible that the electron-deficient hexagonal 
boron network can be stabilized in part or whole through the formation of 3c2e 
bonds with hydrogen, as shown in Fig.~\ref{B2H2-I-struc}, 
which we denote as ``B$_2$H$_2$". The boron atoms in B$_2$H$_2$ form 
borane-like 3c2e bonds with bridging hydrogen atoms and connect 
to their third in-plane neighbor via a conventional
$sp^2$ hybridization (shown in Fig.~\ref{B2H2-I-struc}). 
Note that the hydrogen atoms are bound to the hexagonal 
boron plane in pairs, one above, and the other below 
the plane. 
These weakly bonded hydrogens 
are crucial to stabilize the boron network and are 
desirable for the release of hydrogen at practical conditions.
The extended boron framework in this solid 
boron-hydride structure is quite distinct from a collection of individual 
diborane molecules.  Preservation of the backbone boron structure over cycling 
(by means of kinetic barriers against framework decomposition)
would be extremely advantageous for regeneration as compared to traditional boron hydrides, i.e., boranes.

\section*{Methods: Computational details}
\label{method}

The density functional theory~\cite{kohnsham} based first-principles 
calculations in the present work are performed using 
the QUANTUM ESPRESSO~\cite{ESPRESSO} package. 
The generalized gradient approximation (GGA) of Perdew, Burke, and 
Ernzerhof~\cite{pbe} is used for the exchange and correlation
terms. Ultrasoft pseudopotentials~\cite{vanderbilt} are used 
to describe the electron-ion interactions. 
The plane-wave energy cutoff is set at $50\unit{Ry}$ for
the wave functions expansion and $500\unit{Ry}$ for the charge 
density. All structures are relaxed until the forces 
on each atom are less than $10^{-5}\unit{Ry/a.u.}$ and the stress 
less than $1\unit{GPa}$. 
In this work, we focus on the properties of single-layer B$_2$H$_2$. 
Inter-layer binding will be investigated in the near future using 
a recently developed method for efficient calculations of the van der 
Waals energy~\cite{yiyang}. The charge density for a single-layer
B$_2$H$_2$ is sampled with a $k$-point set of 32$\times$32$\times$8 in the
Brillouin zone. The phonon dispersion is calculated using the
density functional perturbation theory~\cite{Baroni}. The 
dynamical matrices are calculated on a 6$\times$6$\times$4 $q$-point. Next, the 
force constant matrices in real space are obtained by Fourier-transforming
the dynamical matrices. Finally, the dynamical matrix at any $q$-point
is obtained by inverse Fourier-transformation of the force
constant matrices. In the following sections, we discuss
the electronic, structural, and dynamical properties for the proposed 
solid boron hydride B$_2$H$_2$.

\begin{figure}[htb]
\includegraphics[width=0.45\textwidth]{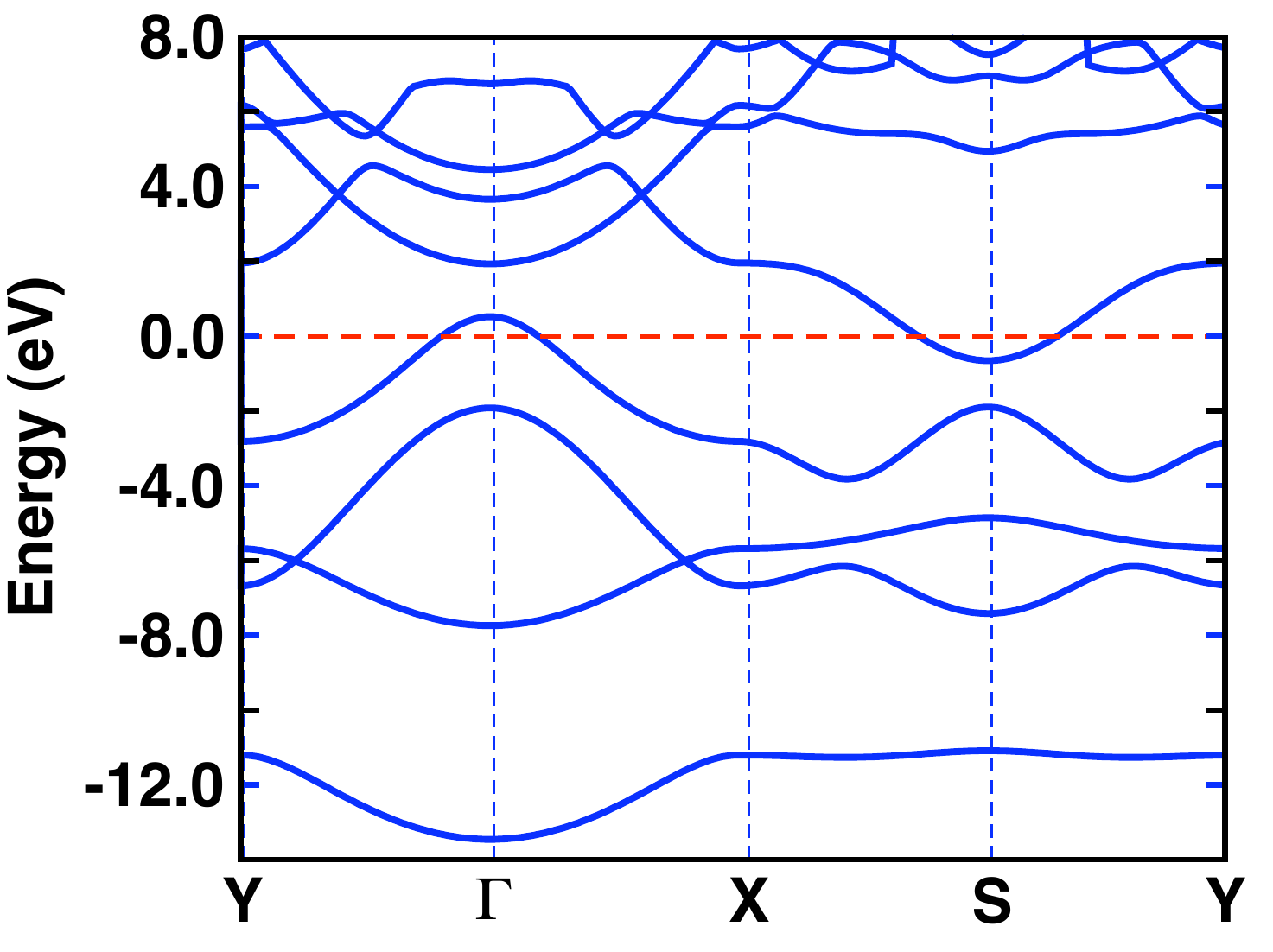}
{\bf \caption{ Band structure of a single-layer of B$_2$H$_2$. The Fermi level is at 0\unit{eV} and shown with red dotted line.}
\label{band-B2H2-I}}
\end{figure}

\begin{figure}[htb]
\includegraphics[width=0.45\textwidth]{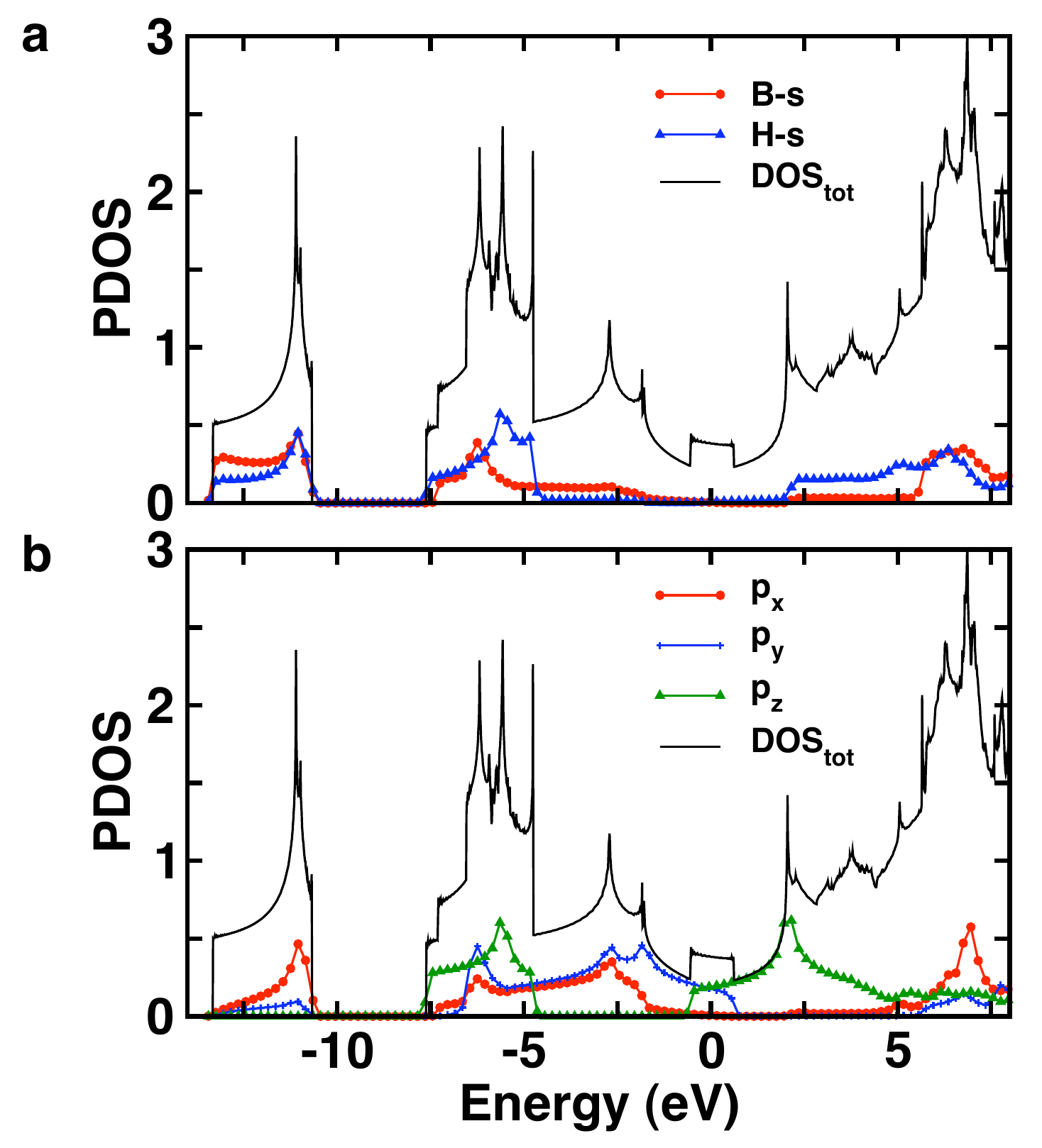}
{\bf \caption{\bf Projected density of states (PDOS) and total density of states (DOS$_{\textrm {tot}}$) in a single-layer B$_2$H$_2$. a, PDOS for B 2s and H 1s, and b, PDOS for B-2p$_x$, B-2p$_y$ and B-2p$_z$. The Fermi level is at 0\unit{eV}. PDOS units are states/eV/cell. }
\label{pdos-B2H2-I}}
\end{figure}

\section*{Structural and electronical properties of single-layer B$_2$H$_2$}
\label{struc_electronical}

The proposed B$_2$H$_2$ layered structure has a $D_{2h}$ symmetry which is
the same as that of diborane.
The calculated bond angles and bond lengths of B$_2$H$_2$ are close to those in diborane. 
The boron$-$boron (B$-$B) distance of the 3c2e bond is $1.82\unit{\AA}$, to
be compared with $1.76\unit{\AA}$ in diborane. The $sp^2$ B$-$B bond length ($1.76\unit{\AA}$)
is substantially shorter than the 3c2e B$-$B bond.
For comparison, the theoretical B$-$B bond length
in MgB$_2$ calculated within the GGA is $1.77\unit{\AA}$. 
The hydrogen-boron (H$-$B) bridge bond length is $1.33\unit{\AA}$. This
is essentially the same as that in diborane and is
substantially longer than the $sp^2$ H$-$B bond ($1.19\unit{\AA}$).
The B$-$H$-$B bridge bond angle is $86^{\textrm{o}}$, to be compared with $83^{\textrm{o}}$ 
in diborane. These hydrogen bridge bonds are crucial to stabilize 
the boron network.

The band structure and projected density of states 
of the single-layer B$_2$H$_2$ are shown in
Fig.~\ref{band-B2H2-I} and Fig.~\ref{pdos-B2H2-I} respectively.
It is interesting to compare planar B$_2$H$_2$ to MgB$_2$. Like MgB$_2$, the system is metallic. However, unlike MgB$_2$, the degeneracy of the $p-\sigma$ bands
at the $\Gamma$-point in MgB$_2$ is now removed due to lowering of the symmetry from $D_{6h}$
to $D_{2h}$. As a result, only one of the $p-\sigma$ bands crosses the Fermi 
level and the hole pocket near the $\Gamma$-point comes from the upper 
$p-\sigma$ band. The degeneracy of the $p-\pi$ bonding and antibonding 
states near the Fermi level is also removed. Similar to that in MgB$_2$,
the Fermi surface of B$_2$H$_2$ also consists of both a $p-\sigma$ derived 
hole pocket and a $p-\pi$ derived electron pockets. 
The projected density of states plot also shows significant involvement
of the hydrogen $s$ states in the bonding.

\section*{Phonon dispersion and dynamical stability}
\label{phonon}

\begin{figure}[htb]
\includegraphics[width=0.45\textwidth]{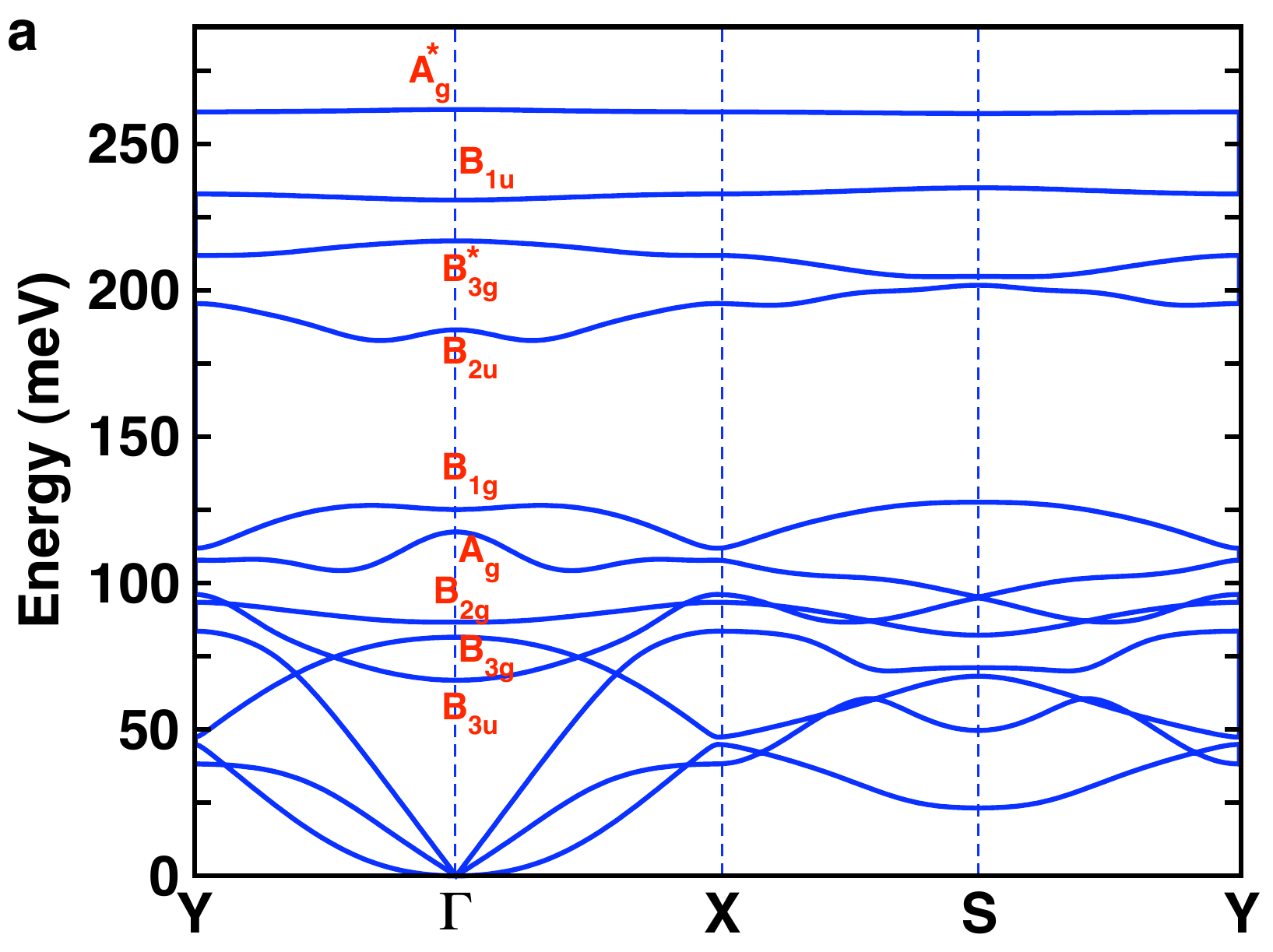}
\includegraphics[width=0.45\textwidth]{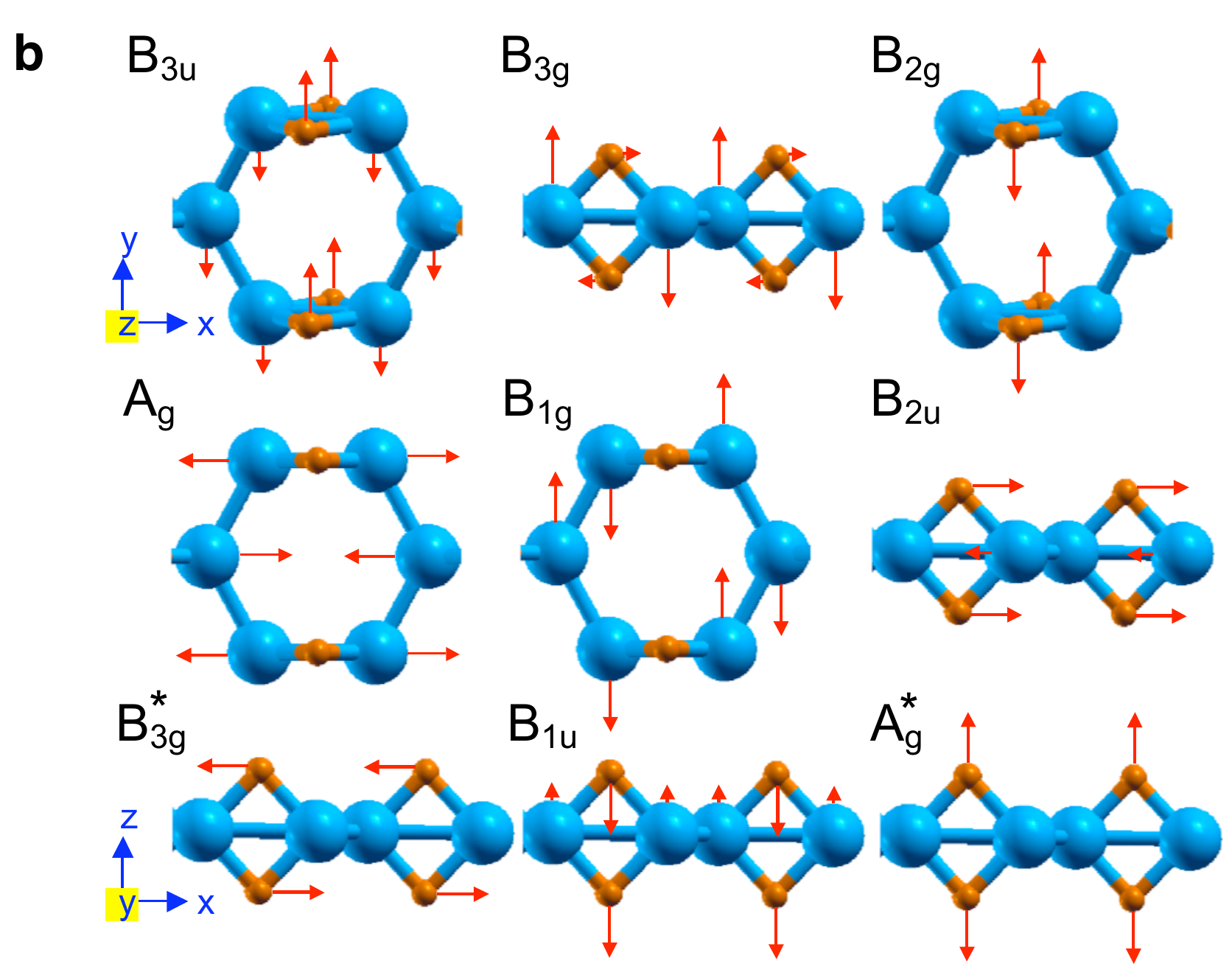}
{\bf \caption{\bf Phonon dispersion: a, The calculated phonon dispersion for single-layer B$_2$H$_2$. The labels at $\Gamma$ are the nine optical phonon modes. b, The corresponding phonon polarization vectors of the optical phonons at $\Gamma$ . Boron is blue and hydrogen is orange.}
\label{phdis-B2H2-I}}
\end{figure}

To investigate the dynamical stability of the proposed structure,
we carry out phonon calculations using density functional perturbation theory~\cite{Baroni}.
The calculated phonon dispersion (shown in Fig.\ \ref{phdis-B2H2-I})
reveals no soft phonon modes (i.e., imaginary phonon energies), thus verifying the 
dynamical stability of the B$_2$H$_2$ structure. 
The nine optical phonon modes can be classified using group theory
analysis:
\begin{eqnarray*}
\Gamma_{vib}&=&
A_g^1[\mathrm{H}\mathrm{-}\mathrm{H}(z)]+
A_g^2[\mathrm{B}\mathrm{-}\mathrm{B}(x)]+ 
B_{1g}[\mathrm{B}\mathrm{-}\mathrm{B}(y)]+ \nonumber \\
& &
B_{2g}[\mathrm{H}\mathrm{-}\mathrm{H}(y)] +
B_{3g}^1[\mathrm{H}\mathrm{-}\mathrm{H}(x)]+
B_{3g}^2[\mathrm{B}\mathrm{-}\mathrm{B}(z)]+\nonumber \\
& &
B_{1u}[\mathrm{H}\mathrm{-}\mathrm{B}(z)]+
B_{2u}[\mathrm{H}\mathrm{-}\mathrm{B}(x)]+
B_{3u}[\mathrm{H}\mathrm{-}\mathrm{B}(y)]. \nonumber \\
& &
\end{eqnarray*}

The two $A_g$ modes will mix so that the polarization vectors
of the eigenmodes with the $A_g$ symmetry 
will involve both hydrogen and boron motions. 
We denote the mode with primarily hydrogen motion as $A_g^*$.
Similarly, the two $B_{3g}$ modes will mix and we denote 
the one with primarily hydrogen motion as $B_{3g}^*$. 
The phonon polarization vectors for the 9 optical phonon modes
at $\Gamma$ are also shown in Fig.\ \ref{phdis-B2H2-I}(b).

The highest energy mode ($A_g^*$) is a symmetric stretching of
H atoms perpendicular to the boron plane (i.e. along the z-direction),
whereas the $B_{1u}$ mode involves H$-$B vibrations in the z-direction.
The next two phonon modes ($B_{3g}^*$ and $B_{2u}$) are H$-$H and H$-$B 
vibrations along the direction of the bridging bonds (x-direction). 
The $B_{1g}$ and $A_g$ modes are similar to the $E_g$ modes in the 
AlB$_2$ structure and involve boron motions in the boron 
plane (x-y plane). Note that the energies of these two modes are
very close to the energy of the $E_g$ phonon in AlB$_2$ but
substantially higher than that in MgB$_2$.
The $B_{2g}$ and $B_{3u}$ modes involve
H$-$H and H$-$B vibrations in the y-direction whereas the low energy $B_{3g}$ mode
consists primarily of boron vibration in the y-direction admixed with hydrogen
motion in the x-direction.

One key aspect of the present work is to assess the feasibility of 
the B$_2$H$_2$ structure for hydrogen storage. This 
requires not only a stable backbone structure that binds hydrogen 
with intermediate strength but also a structure that releases hydrogen 
under practical conditions. In the next section, we investigate the doping
effects on the dynamical properties of B$_2$H$_2$ and discuss their implication for 
dehydrogenation.

\section*{Charge doping effects}
\label{DOPING}

Utilizing B$_2$H$_2$ for hydrogen storage requires not only 
a stable boron framework but also a practical hydrogen release mechanism.
As discussed earlier, a planar hexagonal boron layer is not stable alone due to
electron deficiency. If additional electrons are introduced to the 
B$_2$H$_2$ system (for example, by introducing Li), 
the boron network will be less electron deficient. As a result, the 
hydrogen bridge bonds should weaken with increasing charge doping. 
The hexagonal boron backbone structure, on the other hand,
should remain stable under electron doping.
Therefore, one may be able to control the strength of the hydrogen
bridge bonds through electrochemical charge doping. This, when combined with thermal 
activation, may facilitate dehydrogenation and/or rehydrogenation. 

\begin{figure}[htb!]
\includegraphics[width=0.4\textwidth]{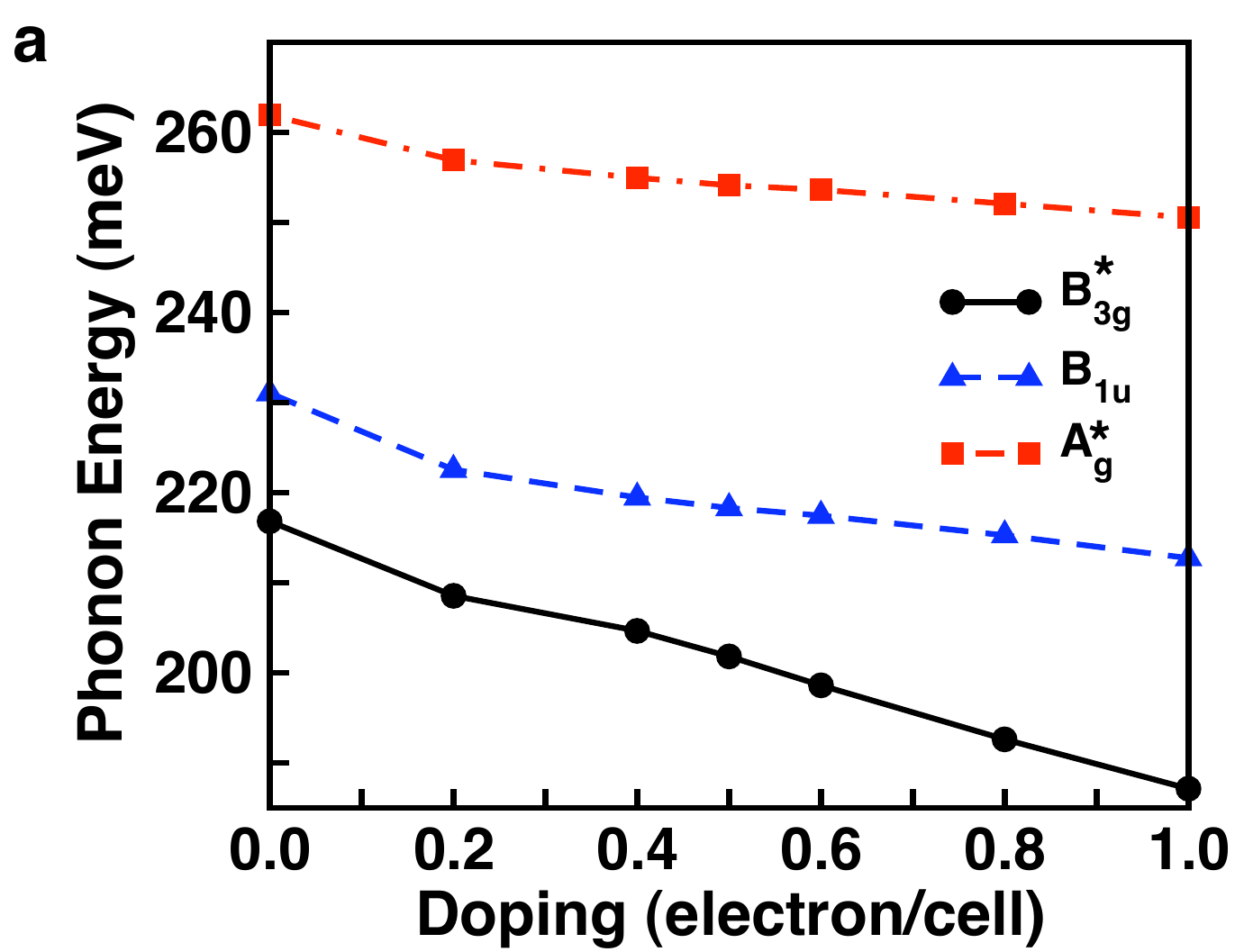}
\includegraphics[width=0.4\textwidth]{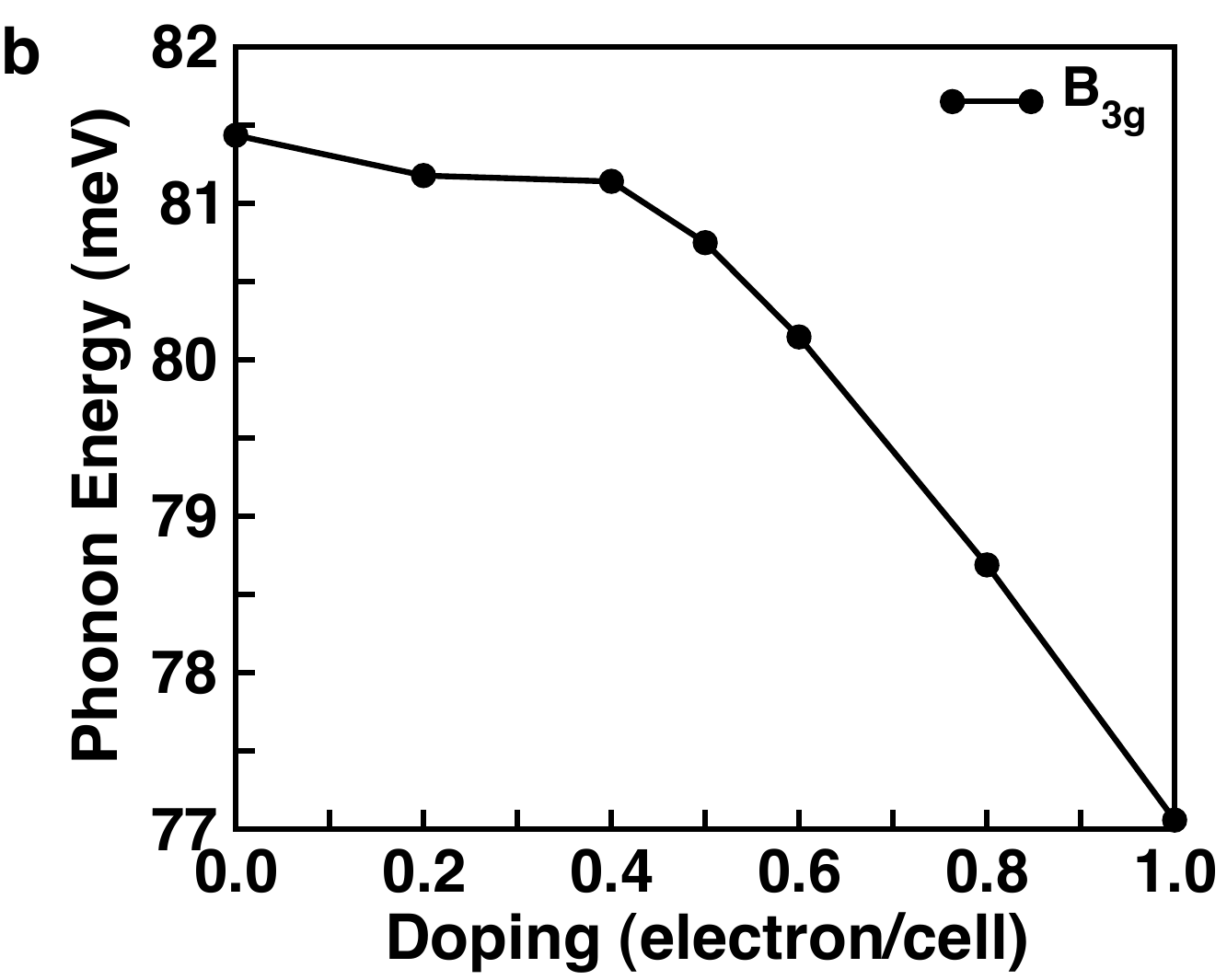}
{\bf \caption{Phonon energy as a function of doping levels for four hydrogen phonon modes at $\Gamma$ for single-layer B$_2$H$_2$ a, the change in the phonon energy for the three higher energy H phonon modes B$^*_{3\textrm{g}}$, B$_{1u}$, and $A_g^*$ and b, for lower energy H phonon mode B$_{3\textrm{g}}$.}
\label{phonon-energy-B3g}}
\end{figure}

Here we investigate this possibility by studying the effects of doping on the zone center phonon energies. 
Our results reveal that some of the phonon modes (especially those involving 
vibrations of hydrogen) are significantly softened by electron doping. 
Figure~\ref{phonon-energy-B3g}(a) and (b) show the softening of the hydrogen-related phonon modes, namely the 
$A_g^*$, $B_{1u}$, $B_{3g}^*$, and $B_{3g}$ phonons. 
The softening of these hydrogen-related phonons is a result of the weakening 
of the hydrogen bridge bonds upon charge doping, which in turn should enhance hydrogen release.

\begin{figure}[htb!]
\includegraphics[width=0.45\textwidth]{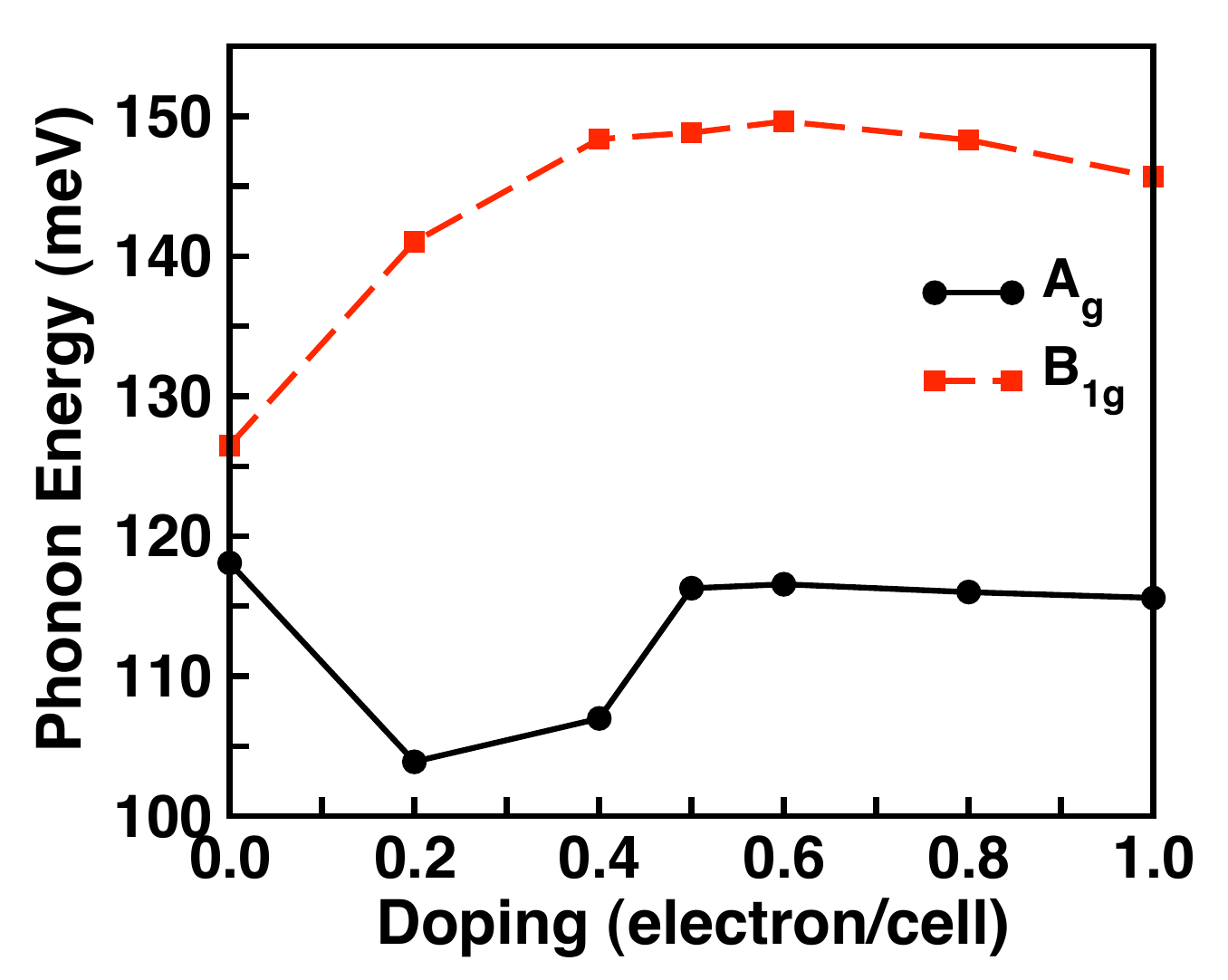}
{\bf \caption{Phonon energy as a function of doping levels for two boron phonon modes: A$_g$ and B$_{1g}$ phonon energies at $\Gamma$ for single-layer B$_2$H$_2$. }}
\label{phonon-energy-G-B}
\end{figure}

In-plane boron modes ($A_g$ and $B_{1g}$)
show more subtle doping dependence. The $B_{1g}$ mode hardens significantly with electron
doping at low doping levels, reflecting a strengthening of the boron network. When doping level
exceeds 0.5 electrons per cell, the $B_{1g}$ mode softens slightly. 
At this doping level, excess electrons start to occupy
high energy antibonding states. Although the system is still dynamically stable,
thermodynamically it may prefer to release hydrogen in exchange for stabilizing the boron network. 
The $A_{1g}$ mode, on the other hand, softens at low doping levels. Since
this mode involves the stretching of both the B$-$B bonds and hydrogen bridge bonds,
the initial softening of this mode shows the combined effects of strengthening 
the boron network and weakening the hydrogen bridge bonds. Clearly,
at small doping, the weakening of the bridge bonds dominate.

\section*{Summary}

In summary, we predict a novel layered solid boron hydride structure (B$_2$H$_2$)
and investigate its structural, electronic and phonon properties 
using density functional based first-principles electronic structure methods.
The structure consists of a graphene-like hexagonal boron network 
and bridge hydrogens which form 3c2e-like multi-center B$-$H bonds.
The formation of multi-center B$-$H bonds is critical to stabilize
the electron-deficient hexagonal boron network.
The phonon dispersion of single-layer B$_2$H$_2$ does not
show any soft phonon modes, thus confirming the dynamical
stability of the proposed structure. We also investigate
the effects of charging on the dynamical properties of B$_2$H$_2$.
The softening of the hydrogen related phonon modes upon
charging suggests a weakening of the hydrogen bridge bonds.
With the presence of extra electrons, the boron network becomes
less electron-deficient therefore reducing the strength of the 
multi-center B$-$H bonds. These results suggest
a charging-assisted hydrogen release mechanism. 
The boron network, however, remains stable upon charging.

\section*{\bf Acknowledgments}
This work was supported by the National Science Foundation
through Grant No. CBET-0844720, and by the Petroleum Research
Fund of the American Chemical Society through Grant No.
48810-DNI10. We acknowledge the computational support provided by the
Center for Computational Research at the University at Buffalo, SUNY.
VHC acknowledges the support from DOE through grant No. DE-FC36-05GO15077.
\section*{\bf Additional information}
The authors declare no competing financial interests.

\end{document}